\documentclass[letter,twocolumn]{jpsj2}
\newcommand{\ket}[1]{\vert #1 \rangle}
\newcommand{\bra}[1]{\langle #1 \vert}

\newcommand{\Figure}[1]{Figure \ref{fig:#1}}
\newcommand{\Fig}[1]{Fig. \ref{fig:#1}}
\newcommand{\eq}[1]{eq. \eqref{eq:#1}}
\newcommand{\Ham}{{\mathcal H}}

\newcommand{\nmax}{{N_{\textsf{m}}}}
\newcommand{\TFP}{T_{\textsf{FP}}}
\newcommand{\deffig}[4]{
  \begin{figure}[htb]
    \begin{center}
      \includegraphics[width=#3 \textwidth]{#2}
    \end{center}
    \caption{ \label{fig:#1} #4}
  \end{figure}
}
\usepackage{color}

\title{Diffusion in the Continuous-Imaginary-Time Quantum World-Line
  Monte Carlo Simulations with Extended Ensembles}
\author{Kenji HARADA and Yuto KUGE}
\inst{
  Graduate School of Informatics, Kyoto University, Kyoto 606-8501
}
\abst{ The dynamics of samples in the continuous-imaginary-time
  quantum world-line Monte Carlo simulations with extended ensembles
  are investigated. In the case of a conventional flat ensemble on
  the one-dimensional quantum $S=1$ bi-quadratic model, the asymmetric
  behavior of Monte Carlo samples appears in the diffusion process in
  the space of the number of vertices. We prove that a local
  diffusivity is asymptotically proportional to the number of vertices,
  and we demonstrate the asymmetric behavior in the flat ensemble
  case. On the basis of the asymptotic form, we propose the weight of an
  optimal ensemble as $1/\sqrt{n}$, where $n$ denotes the number of
  vertices in a sample. It is shown that the asymmetric behavior
  completely vanishes in the case of the proposed ensemble on the
  one-dimensional quantum $S=1$ bi-quadratic model.  }

\kword{diffusion, extended ensemble, quantum Monte Carlo,
  first-passage time, bi-quadratic model}
\begin{document}
\maketitle
% 
% Introduction
% 
Properties of various quantum exotic states and phase transitions
between them have been extensively investigated. For example, quantum
paramagnetic and valence-bond-solid states related to a possible
mechanism to support a novel superconductivity in
cuprates\cite{Anderson1987}, and the existence of
non-Landau-Ginzburg-Wilson type phase transitions between different
broken symmetries\cite{Senthil2004}. In order to numerically investigate such
quantum states and phenomena, unbiased quantum world-line
Monte Carlo (QMC) methods based on a Markov chain are powerful tools,
because they can be applied to large scale systems at low temperatures
and are not limited to the one-dimensional case, if no
negative-sign problem exists. However, even for no negative-sign cases, it is
sometimes difficult to get accurate data from the conventional QMC
simulations, because Monte Carlo samples are trapped near a metastable
state in configuration space. In general, the quality of QMC
simulations deteriorates at low temperatures, because metastable
states are related to degenerate ground states. In fact, the
autocorrelation time of samples in QMC simulations becomes
exponentially large to overcome barriers between metastable states.
In order to eliminate the rapid increase in
autocorrelation times in Monte Carlo simulations, two approaches were
proposed over the last two decades. One is to change the local update method of
samples in a Markov process to a global one that connects metastable
states directly. In fact, the loop algorithm\cite{EvertzLM1993} has
been successful in studies of quantum magnetic phases due to the
global update whose shape is like a loop and which corresponds to a
magnetic correlated domain. However, in some cases using the loop
algorithm, we encountered rapid increases in autocorrelation
time: For example, valence-bond-solid states that break spatial
symmetries on the two and quasi-one dimensional
lattice\cite{HaradaKT2003, Harada2007}. For such cases, the second
approach is probably effective, in which a canonical ensemble is
replaced by an artificial extended ensemble such that Monte Carlo
samples would not be trapped near metastable states. For classical
systems, the extended weight is adjusted such that the appearance ratio
of energy $E$ samples would be flat. They are not trapped near a
metastable state because Monte Carlo samples diffuse in a wide energy
range. Therefore, these methods have been extensively used for studies
of spin glasses and frustrated classical models. However, for quantum
models, they have been tested only in a few cases\cite{Troyer2003,
  Wessel2007}. In order to get conclusive numerical results for
quantum strong correlated phenomena as mentioned above, we need to
understand the property of QMC methods with extended ensembles, and it
is important to improve their efficiencies. In this letter, we will
concentrate the diffusive behavior of samples in the
continuous-imaginary-time QMC simulations with an extended
ensemble. We will report the asymmetric behavior in the diffusion
process for the one-dimensional quantum $S=1$ bi-quadratic (BQ) model
case.  On the basis of our proven asymptotic form of the local
diffusivity of samples, we will propose an optimal ensemble. The
performance of the proposed ensemble will be shown for the case of the
one-dimensional BQ model.

% 
% EEQMC
% 
Although the first formulation of QMC methods with extended ensembles
(EEQMC) was based on high-temperature series expansion
\cite{Troyer2003}, it can be also done on the path-integral
representation with a continuous-imaginary-time limit (see \S\,2.15 in
ref. \citen{Kawashima2004}). In order to provide a brief description
of the EEQMC algorithms on the path-integral representation, we first
consider the definition of the exponential operator:
\begin{equation}
  \label{eq:exp}
  \exp\left(-\beta\sum_b\Ham_b\right)
  =\lim_{M\to \infty}
  \left[\prod_b\left(1-\frac{\beta\Ham_b}{M}\right)\right]^M.
\end{equation}
Inserting the identity operator $1=\sum_{\alpha}
\ket{\alpha}\bra{\alpha}$ with a complete orthonormal basis
$\{\ket{\alpha}\}$ between two adjacent factors in the right-hand side
of \eq{exp}, we obtain the discrete-imaginary-time path-integral
representation of a partition function as
\begin{equation}
  \label{eq:partition_function}
  Z\approx\sum_{\{\ket{S_b(t)}\}}\prod_{t=1}^{M}\prod_{b=1}^{K}\bra{S_{b+1}(t)}\left(1-
    \beta\Ham_b\Delta
  \right)\ket{S_b(t)},
\end{equation}
where $\Ham_b$ is the $b$-th interaction Hamiltonian, $\beta$ is an
inverse temperature, $\Delta=1/M$, $S_{K+1}(t) \equiv S_1(t+1)$,
and $S_1(M+1) \equiv S_1(1)$. Next, we introduce new auxiliary
variables $G_b(t)$ called \textit{graph variables} as
$(1-\beta\Ham_b\Delta)=\sum_{G_b(t)=0,1}(-\beta\Ham_b\Delta)^{G_b(t)}$.
Then, \eq{partition_function} is rewritten as
\begin{align}
  \label{eq:partition_function3}
  &Z \approx \sum_{S}\sum_{G}\beta^{n(G)} W_0(S,G)
  =\sum_n \beta^n \Omega(n),\\
  \label{eq:weight0}
  &W_0(S,G)=\prod_u\bra{S_{u'}}\left(-\Ham_u\Delta\right)^{G_u}\ket{S_u},\\
  \label{eq:omega}
  &\Omega(n)=\sum_{S}\sum_{(G|n(G)=n)} W_0(S,G),
\end{align}
where $u$, $u'$, $S$, and $G$ denote $(b,t)$, $(b+1,t)$, $\{S_u\}$, and
$\{G_u\}$, respectively, and $n(G)=\sum_u G_u$. In the following, a
graph variable $G_u$ that takes a value $1$ is called a \textit{vertex} and the
number of vertices $n(G)$ is called a \textit{vertex number}. This
representation is the mathematical background required to describe the remarkable QMC
algorithms that have been developed during the last two
decades\cite{Kawashima2004}.  In particular, we can take a limit of a
continuous-imaginary time, $\Delta \to 0$, in the level of QMC
algorithms (see \S\,2.5 in ref. \citen{Kawashima2004} for
details). In the remainder of this paper, we consider the EEQMC
algorithm on the continuous-imaginary time, because it has no
systematic error from the discretization of an imaginary time.

The vertex number $n(G)$ in a canonical ensemble statistically
corresponds to an inverse temperature $\beta$ because $\langle n(G)
\rangle_\beta = \beta \langle -\Ham \rangle_\beta$, where $\langle
\cdot \rangle_\beta$ is a canonical ensemble average. Therefore, if we
adjust the weight of a configuration $(S,G)$ so that the frequency of
obtaining the vertex number $n$ would be independent of $n$, i.e.,
flat\cite{Troyer2003}, we can sample various configurations in a wide
inverse temperature range. From \eq{partition_function3}, $\Omega(n)$
is regarded as the density of states with a fixed vertex number
$n$. Hence, if the factor $\beta^{n(G)}$ in \eq{partition_function3}
is replaced by $1/\Omega(n(G))$, it can be done. In general, in order
that the appearance ratio of configurations with a vertex number $n$
is $P_v(n)$, the extended ensemble weight of a configuration $(S,G)$
has to be ${P_v(n(G))W_0(S,G)}/{\Omega(n(G))}$. However, we need to
guess $\Omega(n)$ from the QMC samples themselves, because $\Omega(n)$ is
not known a priori. Fortunately, some sophisticated methods have been
proposed.  \cite{Wang2001, Oliveira1998} In our simulations, we have
used the broad-histogram relation for the vertex
number\cite{Yamaguchi2004} as
\begin{equation}
  \label{eq:BHR}
     \frac{\Omega(n+1)}{\Omega(n)}
     =\frac{\left\langle{\textsf{diag}}(-\Ham)\right\rangle_n}
     {n+1-\langle n_K\rangle_{n+1}},
\end{equation}
where $\langle Q \rangle_n$ denotes the micro-canonical ensemble
average of an operator $Q$ at a fixed vertex number $n$,
${\textsf{diag}}(Q)$ refers to the diagonal part of an operator $Q$, and
$n_K$ is the number of kinks at which a state changes, i.e., $\langle
S_{u'} \vert S_{u}\rangle=0$. Because the right-hand side in \eq{BHR}
can be directly estimated in the EEQMC simulations, $\Omega(n)$ can be
calculated from this recursion formula.  We should note that this
estimation method is independent of the dynamics of the EEQMC samples;
it is not based on the histogram of the appearance of a vertex
number in the EEQMC simulations.

% 
% Diffusion
% 
However, after $\Omega(n)$ is sufficiently estimated, the dynamics of
the EEQMC samples do not seem to be a regular random walk in the vertex number
space. In particular, in order to investigate this behavior, we focus
on the first-passage time (FPT) of EEQMC samples regarded as random
walkers in a vertex number space. The FPT is defined as the time at
which a random walker first reaches a threshold value. Because the
movement of an EEQMC sample is usually restricted to the interval
$[N_a,N_b]$ in the vertex number space, two types of FPTs are defined
as a QMC sample moves to $N_b$ from $N_a$, and vice versa. In the
following, the former is called \textit{forward} and the latter is
called \textit{backward}.
\deffig{FPT}{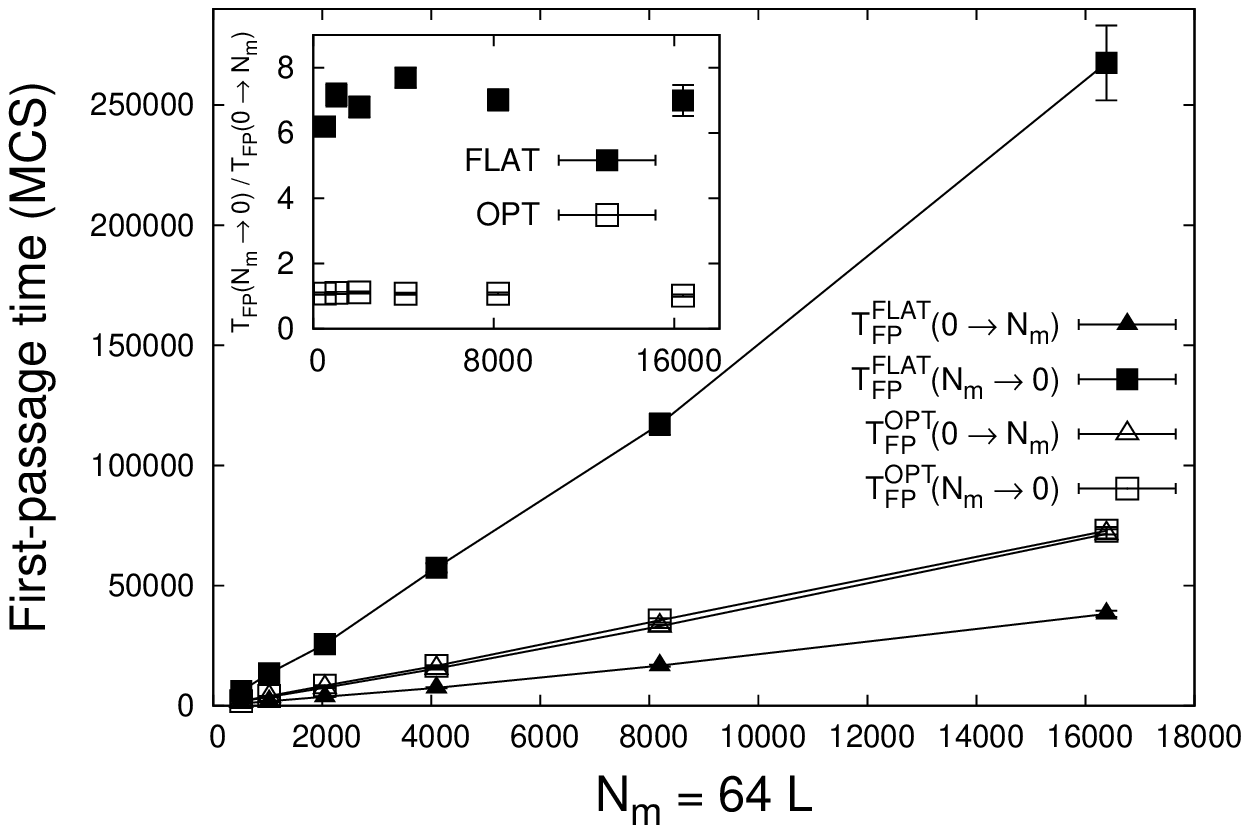}{0.45} {First-passage times (FPT) of
  EEQMC samples for the one-dimensional quantum $S=1$ bi-quadratic
  model. Filled triangle and square symbols are forward and backward
  FPTs for a flat ensemble, respectively. Triangle and square symbols
  are forward and backward FPTs for the ensemble defined by
  \eq{OPTQMC}, respectively. In the inset, the ratios of backward and
  forward FPTs are shown.}

\Figure{FPT} shows forward and backward FPTs, $\TFP(0 \to \nmax)$ and
$\TFP(\nmax \to 0)$, for the one-dimensional BQ model with chain
lengths $L=8, 16, 32, 64 ,128,$ and $256$. The Hamiltonian of the BQ model
on the one-dimensional lattice is $\Ham=-\sum_{i} (\textbf{S}_i \cdot
\textbf{S}_{i+1})^2$, where $\textbf{S}_i$ denotes an $S=1$ spin
operator at the $i$-th site.  Because this model is integrable, the
ground state is well known as the dimerized state in which singlet
pairs align and completely cover the one-dimensional lattice. This
dimerized state is twofold degenerate and spontaneously breaks a
translational symmetry. And the autocorrelation time in QMC
simulations is the round-trip time between such degenerate states. As
mentioned above, even if we use a loop algorithm for the BQ
model\cite{Harada2001}, it grows rapidly at low
temperatures. Therefore, this model is suited to the test of the EEQMC
algorithms. In all cases, $\nmax$ is proportional to the chain length
$L$ as $\nmax = 64 L$. It corresponds to a constant low
temperature enough to calculate a ground state, because the one-dimensional
BQ model has a gap. The total number of EEQMC samples for a chain
length is sufficiently large for estimating FPTs: For example, $32$ independent
runs with $6 \times 10^6$ Monte Carlo sweeps (MCSs) for $L=256$.  We
should note that MCS is adopted as a unit of time. One MCS for a
configuration $(S,G)$ in the continuous-imaginary-time EEQMC algorithm
consists of three steps: (i) deciding a new vertex number $n'$ under a
given $S$-configuration, (ii) assigning new graphs $G'$ with the
vertex number $n'$ to a given $S$-configuration, and (iii) choosing a new
$S'$-configuration under the given graphs $G'$.  It is possible to
measure observables only at the time that these three steps are completed.
In \Fig{FPT}, forward and backward FPTs for a flat ensemble
increase almost linearly, but the backward FPT is always larger than
the forward one: For example, the ratio of two FPTs is 7.0(5) for
$L=256$ (see the inset of \Fig{FPT}). Thus, the EEQMC samples for a flat
ensemble move quickly from high temperatures to low ones, but slowly
in the reverse direction. In order to improve the efficiency of the EEQMC
algorithms, it is necessary to correct this asymmetric behavior.

% 
% Local Diffusivity
% 
When we make a new $G'$ configuration under a fixed $S$-configuration,
the vertex ($G_u=1$) at a kink can not be removed, because the local
weights in $W_0(S,G)$ at a kink becomes zero: $\langle S_{u'} \vert
S_{u}\rangle=0$. Therefore, if the number of kinks is $n_K$, the
probability $p(n'|n_K)$ to choose the next vertex number $n'$ in step
(i) is proportional to the sum of weights of the configurations that
have unchanged $n_K$ vertices at kinks and new inserted $(n'-n_K)$
ones into a given $S$-configuration:
\begin{equation}
  \label{eq:cond_prob}
  p(n'|n_K)\propto 
  \frac{P_v(n')\left[\sum_{(u|G_u=0)}\bra{S_{u'}}
      \left(-\Ham_u\Delta\right)
      \ket{S_u}\right]^{(n'-n_K)}}
  {\Omega(n')(n'-n_K)!}.
\end{equation}
For finite-size systems, if the vertex number is sufficiently large, the
right-hand side in \eq{BHR} is converged. Using limiting values as
$w_0=\lim_{n\to\infty}\left\langle {\textsf{diag}}(-\Ham)
\right\rangle_n$ and $r_K=\lim_{n\to\infty}\left\langle
  n_K\right\rangle_n/n$, the asymptotic form of $\Omega(n)$ is as
\begin{equation}
  \label{eq:asym_omega}
  \Omega(n) \approx \frac{\Omega(n-1)}{n} \left(\frac{w_0}{1-r_K}\right)
  \propto
  \frac{1}{n!}\left(\frac{w_0}{1-r_K}\right)^n.
\end{equation}
Substituting \eq{asym_omega} into \eq{cond_prob}, we find that the
main factor of $p(n'|n_K)$ is the negative binomial
distribution. Using the limit theorem for the negative binomial
distribution, we obtain the asymptotic form of $p(n'|n_K) \propto
P_v(n')\exp[-(n'-m_0)^2/2(m_0/r_K)] $, where $m_0 \equiv (r_K^{-1} -
1)n_K$. Here, we assume that $P_v(n')$ is a slowly varying function.
And if we assume that the probability $p(n_K|n)$ that the number of
kinks in a configuration with a vertex number $n$ is $n_K$ is
equivalent to the probability $p(n|n_K)$, the probability to choose the
next vertex number $n'$ from configurations with a vertex number $n$
is
\begin{align}
  p(n\to n') &= \int_0^n dn_K~p(n'|n_K)p(n_K|n), \\
  \label{eq:jump}
  &\approx
  N_{\textsf{G}}\left(n,\left[\frac{2(1-r_K)}{r_K}\right]n\right),
\end{align}
where $N_{\textsf{G}}(m,\sigma^2)$ denotes the Gaussian distribution
with mean $m$ and variance $\sigma^2$. Thus, the EEQMC samples almost seem
to be random walkers in the vertex number space, but the local diffusivity
$D(n)$ increases linearly as
\begin{equation}
  \label{eq:LD}
  D(n) \approx \left[\frac{(1-r_K)}{r_K}\right]n,
\end{equation}
where $n$ is the number of vertices in a configuration. \Figure{LD}
shows the local diffusivity $D(n)$ for a flat ensemble ($P_v(n)=1$) in
the EEQMC simulations of the one-dimensional BQ model. The chain length
$L$ is $128$ and the total number of MCSs is $1.05 \times 10^8$. The
local diffusivity in \Fig{LD} is approximately linear in the region
above the vertex number 100 (see also the left-top inset of \Fig{LD}).
The solid line in \Fig{LD} is a linear function predicted in \eq{LD}
with $r_K=0.6108(1)$, which is evaluated from the QMC simulations. The
predicted line is consistent with the local diffusivity in the region
above the vertex number 1000. And the discrepancy between them is
never more than $6\%$ at all vertex numbers below 1000 but zero. Next,
we checked the dependence of the local diffusivity on extended
ensembles. In the right-bottom inset of \Fig{LD}, the ratios between local
diffusivities in two different ensembles are shown.  Because these
values are approximately equal to one, the local diffusivity is almost
unaffected by the choice of extended ensembles. For other system-size
cases, the same results were obtained. Therefore, the local diffusivity of
the EEQMC samples is described well by \eq{LD}.
\deffig{LD}{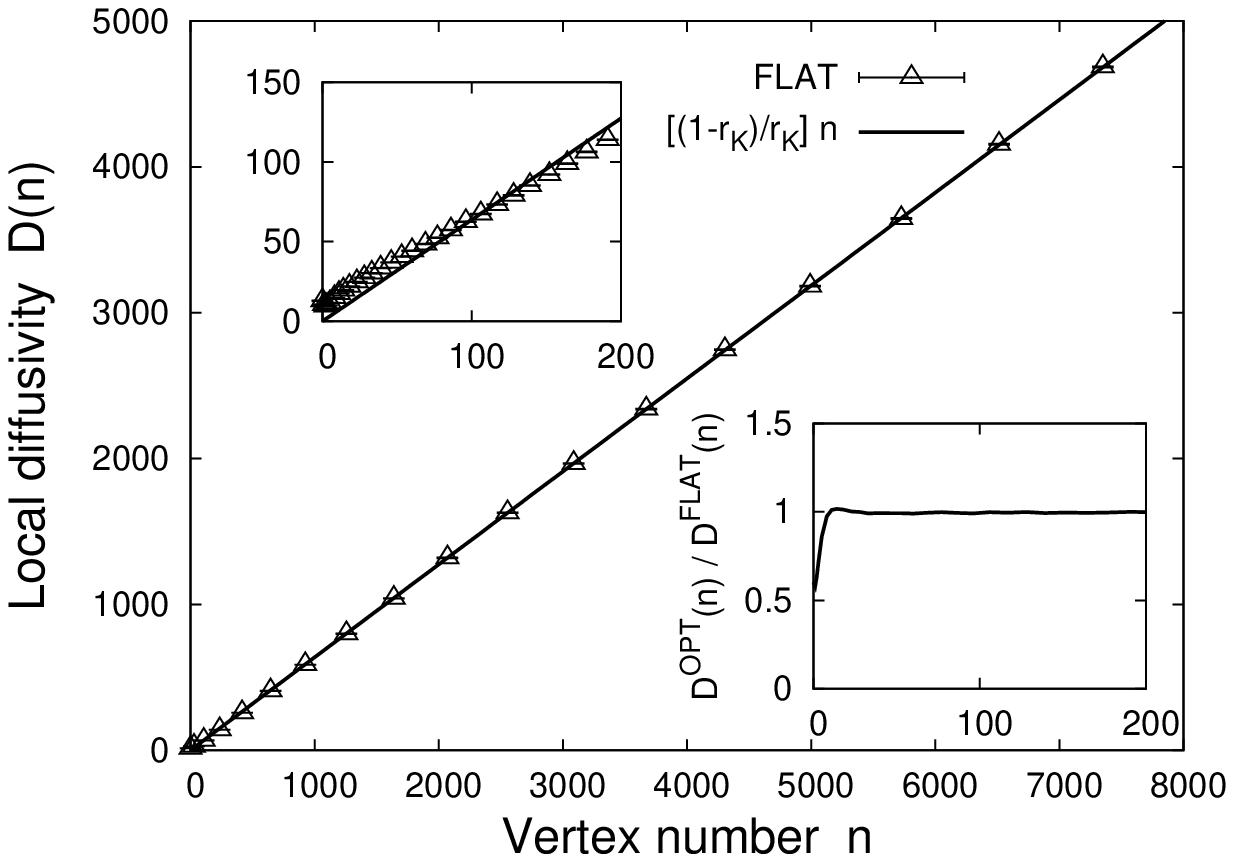}{0.5} {Local diffusivity in the EEQMC simulations
  of the one-dimensional quantum $S=1$ bi-quadratic model. The chain
  length $L$ is $128$. The extended ensemble is flat: $P_v(n)=1$. The
  solid line shows the predicted linear increase in \eq{LD}. The value
  of $r_K$ is $0.6108(1)$, which is evaluated from the QMC simulations. In
  the left-top inset, the local diffusivity in the small vertex number
  region is shown. In the right-bottom inset, the ratios between local
  diffusivities in a flat ensemble and the optimal one defined by
  \eq{OPTQMC} are shown.}

From \eq{jump}, the behavior of the samples in the EEQMC simulations may be
described well by a Fokker-Planck equation (FPE) on a vertex number
range $[N_a,N_b]$\cite{Nadler2007}. Using the theory of first-passage
processes\cite{Redner2001}, we can explicitly obtain the first-passage
times for a one-dimensional FPE as
\begin{align}
  \label{eq:FPT11}
    \TFP(N_a \to n)&=\int_{N_a}^n dx P_v(x)\int_x^n \frac{dx'}{D(x')P_v(x')},\\
  \label{eq:FPT12}
    \TFP(N_b \to n)&=\int_n^{N_b} dx P_v(x)\int_n^x \frac{dx'}{D(x')P_v(x')}.
\end{align}
If we assume the linear increase of the local diffusivity $D(n)$ as in
\eq{LD}, FPTs for a flat ensemble ($P_v(n)=1$) are as
\begin{equation}
  \label{eq:RatioFPT}
    \TFP^{\textsf{FLAT}}(0\to\nmax)\approx\frac{\nmax}{r_K^{-1}-1},
    \ 
    \frac{\TFP^{\textsf{FLAT}}(\nmax\to 0)}{\TFP^{\textsf{FLAT}}(0\to\nmax)} \approx \log \nmax.
\end{equation}
These results are qualitatively consistent with the behavior of FPTs
in \Fig{FPT}. The forward FPT increases linearly and the backward one
is always larger than the forward one and their ratio varies
slowly. Thus, the qualitative behavior of the EEQMC samples is described
well by the FPE with \eq{LD}.

% 
% Optimal ensemble
% 

In order to correct the asymmetry between forward and backward FPTs
for a flat ensemble, an extended ensemble $P_v(n)$ should be adjusted
so that the right-hand side in \eq{FPT11} would be equivalent to that
in \eq{FPT12}. While it typically cannot be uniquely determined, a
special solution exists: $P_v(n)=1/\sqrt{D(n)}$. This special ensemble
was first derived from a maximization of random walker
flows\cite{Trebst2004}. Using this ensemble, we find that the forward
FPT is not only equivalent to the backward one, but also the total FPT, $
\TFP(0\to\nmax) + \TFP(\nmax\to 0)$, is
minimized\cite{Nadler2007}. Therefore, from \eq{LD}, the optimal
ensemble for the EEQMC methods is
\begin{equation}
  \label{eq:OPTQMC}
  P_v^{\textsf{OPT}}(n) = \frac{1}{\sqrt{D(n)}} \propto \frac{1}{\sqrt{n}}.
\end{equation}
Substituting eqs. \eqref{eq:LD} and \eqref{eq:OPTQMC} into
eqs. \eqref{eq:FPT11} and \eqref{eq:FPT12}, we find that the forward and
backward FPTs for the optimal ensemble are equivalent and become
twice as large as the forward one for a flat ensemble:
\begin{equation}
  \label{eq:OPTFPT}
  \TFP^{\textsf{OPT}}(\nmax\to 0)=\TFP^{\textsf{OPT}}(0\to\nmax)=2 \TFP^{\textsf{FLAT}}(0 \to \nmax).
\end{equation}
The forward and backward FPTs in the EEQMC simulations with the optimal
ensemble are shown in \Fig{FPT}.  Here, although \eq{LD} is the
asymptotic form, we use \eq{OPTQMC} for all vertex numbers except zero,
and the value at the vertex number zero is defined by that at the
vertex number one: $P_v(0) \equiv P_v(1)$.  In \Fig{FPT}, the forward
FPT for our proposed ensemble is consistent with the backward one (see
also the inset of \Fig{FPT}). And, as in \eq{OPTFPT}, they are
approximately twice as large as the forward FPT for a flat
ensemble. Thus, the asymmetry is completely corrected by the ensemble
in \eq{OPTQMC}.

As mentioned above, we adopt MCS as the unit of time. But the actual
computational times for one MCS are not constant. In fact, in many
cases the number of local steps in one MCS is proportional to the
vertex number in a sample. Therefore, the local step can be adopted as
the unit of time. In this case, from \eq{LD}, we find that the local
diffusivity in units of local steps is independent of the vertex
number. Hence, the optimal ensemble in units of local steps is flat. In
other words, the optimal ensemble for computational times is as
$P_v^{\textsf{LOPT}}(n)=1/n$. The total local steps of FPTs is
$\frac{3}{4}$ times that for $P_v^{\textsf{OPT}}$. But the number
of samples in the region of large vertex numbers decreases more than
that for $P_v^{\textsf{OPT}}$. In order to calculate a canonical
ensemble average of an observable at an inverse temperature $\beta$,
it is necessary to calculate the reweighted summation of
micro-canonical ensemble averages. Vertex numbers that mainly
contribute to it are in the region of which the width is proportional
to $\sqrt{n(\beta)}$, where $n(\beta)$ is a center of the
region. Therefore, in the case of $P_v^{\textsf{OPT}}$, the number of
samples that contribute to a canonical ensemble average is constant
at all inverse temperatures. But that for the ensemble
$P_V^{\textsf{LOPT}}$ decreases at low temperatures.

%
% Conclusion
%
In summary, we considered the diffusion of samples in the
continuous-imaginary-time EEQMC simulations. In particular, the
asymmetric behavior of FPTs of EEQMC samples was reported in
detail. We proved that the local diffusivity of the EEQMC samples is
asymptotically proportional to the vertex number. And it was shown
that the asymptotic form is consistent with the local diffusivity in
the EEQMC simulations of the one-dimensional BQ model in the wide
region of the vertex numbers. Using this result and the theory of
first-passage processes, we demonstrated the asymmetric behavior of
FPTs for a flat ensemble case and proposed an optimal ensemble for
the continuous-imaginary-time EEQMC simulations in order to correct
the asymmetric behavior. It was shown that the asymmetric behavior on
the one-dimensional BQ model completely vanishes in the case of the
proposed ensemble.

% 
% Acknowledgments
% 
The author would like to thank Naoki Kawashima for useful comments.
The computation in the present work is executed on computers at the
Supercomputer Center, Institute for Solid State Physics, University of
Tokyo.  The present work is financially supported by MEXT Grant-in-Aid
for Scientific Research Wakate (B) 19740237 (2007) and Kiban (B)
19340109 (2007) and by Next Generation Supercomputing Project,
Nanoscience Program, MEXT, Japan.

% 
% References
% 

\end{document}